# ARABIC TEXT WATERMARKING : A REVIEW

Reem Ahmed Alotaibi[1] AND Lamiaa A. Elrefaei[1,2]

[1]Computer Science Department, Faculty of Computing & Information Technology, King Abdulaziz University, Jeddah, Saudi Arabia
[2]Electrical Engineering Department, Faculty of Engineering, Shoubra Benha University, Cairo, Egypt

***ABSTRACT***

*The using of the internet with its technologies and applications have been increased rapidly. So, protecting the text from illegal use is too needed . Text watermarking is used for this purpose. Arabic text has many characteristics such existing of diacritics , kashida (extension character) and points above or under its letters .Each of Arabic letters can take different shapes with different Unicode. These characteristics are utilized in the watermarking process. In this paper, several methods are discussed in the area of Arabic text watermarking with its advantages and disadvantages .Comparison of these methods is done in term of capacity, robustness and Imperceptibility.*

## 1. INTRODUCTION

Digital watermarking is one of the most important techniques uses information hiding in order to protect any kind of media from the possible kinds of attacks. Digital watermarking is defined in[1] as " a process that embeds or inserts extra information, named the watermark or mark, into the original data to generate the output which is called a watermarked or marked data".

The watermark system goes through three stages[2] :firstly the process of generation and embedding the watermark in the original media. Then, possible attacks could occur in the transmission of the signal through the watermark channel. Finally, the process of detection the embedded watermark.

Most of watermarking techniques can be classified into several categories based on some criteria [3] as shown in figure 1. They can be visible or invisible. In visible watermarking technique ,the watermark can be observed by the human eye like logos. Invisible watermarking technique are used more in Steganography. They can also be classified into robust , fragile and semi-fragile techniques based on the modification could be occur on the watermark. In a robust watermarking technique, the watermark cannot be affected from attacks when it detected or extracted. The watermarking technique is called a fragile technique, if the watermark is changed or destroyed. In semi-fragile watermarking technique, the watermark can be affected from some types of attacks. Watermarking techniques can be blind or non-blind. A blind technique does not require the original data for detection or extraction. While in non-blind technique ,the original data is needed for detection or extraction process.

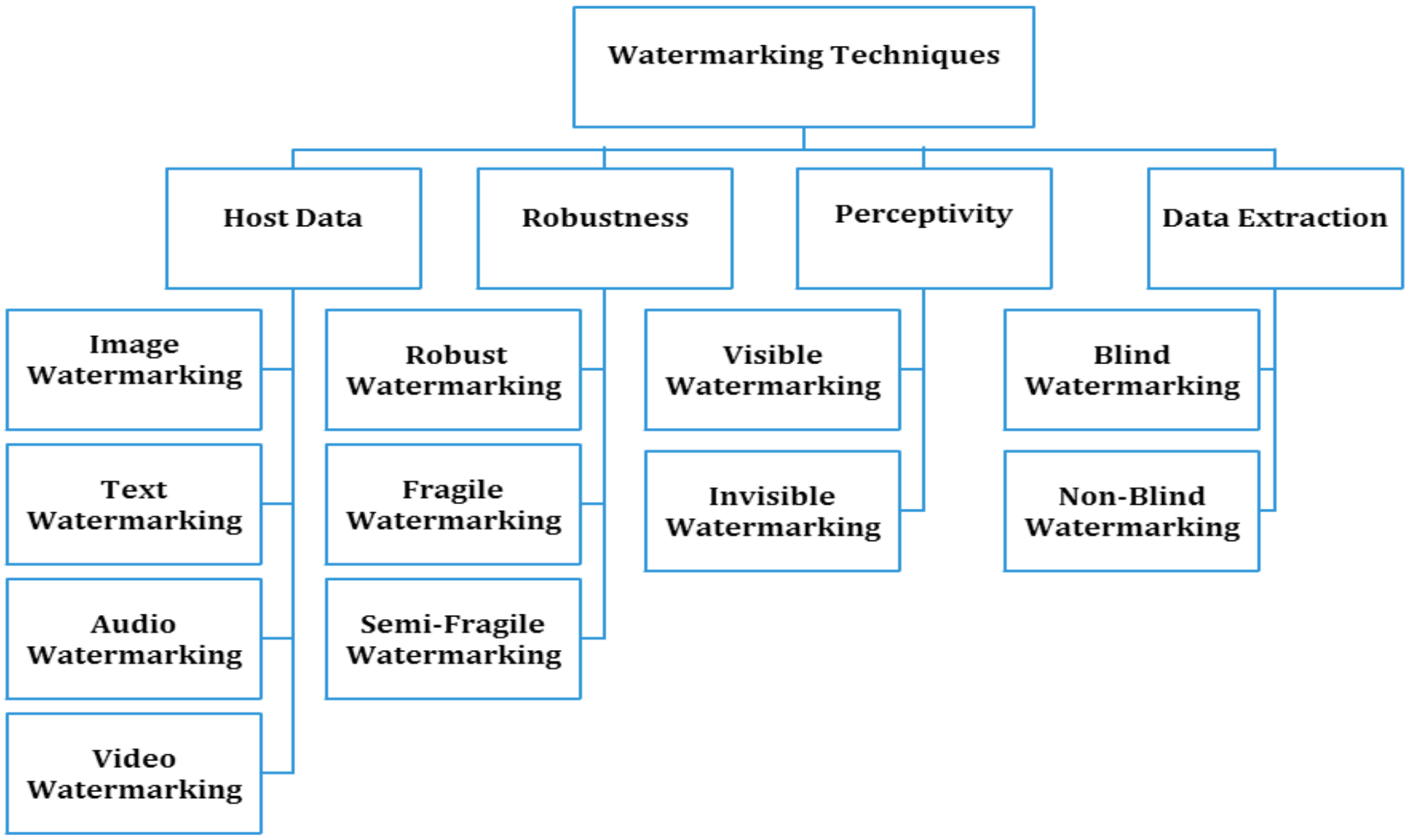

Figure 1.Watermarking types based on some criteria.

An effective watermarking technique should have several requirements[4][5]. These requirements differ depending on the application and the type of data to be watermarked. The watermarking requirements in general are:

- **Robustness:** the watermarking method should resist any kind of distortion or altering by malicious data processing or any kind of attack. The watermark should still be detectable in the extraction process[6].
- **Security:** the watermark must remain hidden from unauthorized detection. This property is achieved by using cryptographic keys.
- **Capacity :**which means the number of hidden bits the technique can encode in a unit amount of time. The watermark should convey a sufficient amount of information. The capacity ratio in text watermarking could be computed as:
  Capacity ratio(Bits/KB) =(Watermark data (Bits))/(Cover file (KB)). [7][8][9-13]
  Capacity ratio (%) =(File size(Bytes))/(Cover size(Bytes) )×100 .[8][14][15][16]
- **Imperceptibility:** the original text should not noticeably destroyed after the watermark embedding process , the modifications are done by a small amount.
- 

It is preferable that text watermarking should be easy to implement, imperceptible, robust, and adaptable to different text formats, have high information carrying capacity[17].

The rest of this paper is organized as follows. Section 2 discusses the Arabic text features. In section 3, several Arabic text watermarking techniques are discussed, and section 4 concludes the paper.

## 2. ARABIC TEXT FEATURES

The Arabic Alphabet consists of 28 characters. It has many characteristics for example, the Arabic script is written from right to left and has no equivalent to capital letters unlike English script. The Arabic word could be consisting of completely connected letters such as: مكة,محمد,جبل or a single word may consists of more than one components like:مازن,سارة,جمال . The letters are connected from the horizontal baseline of the word. They have different shapes based on its position in the word or sub-word except Hamza (ء). Each letter can contain from one to four shapes corresponding to four positions: isolated ;the letter is not connected to any other letter, Initial; the letter is connected to the following letter but not to the previous one, middle; the letter is connected to both the previous and following letters, and finally, final; the letter is connected to the previous letter but not to the following one. As we mention above Hamza {ء} takes only one shape which means it cannot linked to any letter. The letters { ا,و,ر,ز,د,ذ} take two shapes: isolated and final and these are the characters that cannot connect to the following letterers. The existing of these letters in the word indicates that the word divided in one or more sub-words. The rest of letters take the four shapes which means they linked to the pervious and the following letters[18][19] .Table 1 shows examples of the three cases.

Table 1. Different Shapes of some Arabic letters.

| Name | Isolated | Initial | Medial | Final |
|---|---|---|---|---|
| Hamza | ء | | | |
| alif | ا | | | ـا |
| Ayn | ع | عـ | ـعـ | ـع |

The Arabic letters classified into pointed and un-pointed as shown in figure 2 . There are similarity of some letters shapes but they differ in the number and placement of points on the letters[18] as in figure 3 .

| Pointed letters | Un-pointed letters |
|---|---|
| ش ز ذ خ ج ث ت ب<br>ي ن ق ف غ ظ ض ة | ص س ر د ح ا<br>و ه م ل ك ع ط |

Figure 2. Pointed and un-pointed Arabic letters.

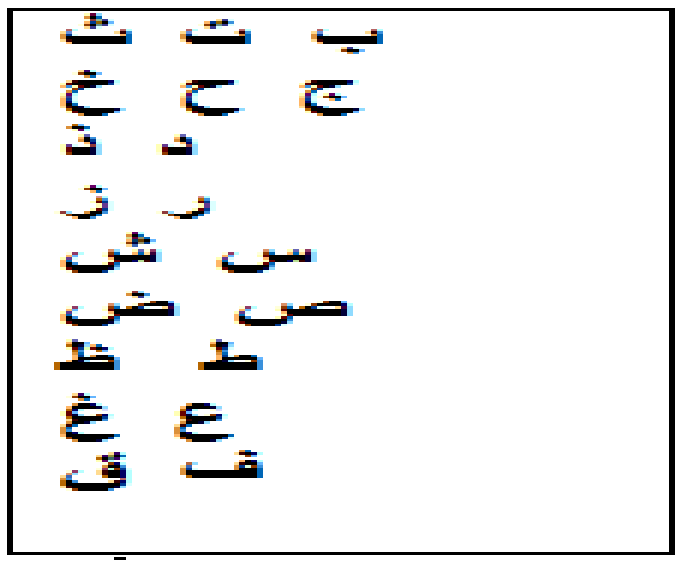

Figure 3. Similar Arabic letters.

Arabic extension character "kashida " is used to expand the space between connected letterers. The kashida refers to a character representing this elongation (-) which increases the length of a line of text. It cannot be added at the beginning or ending of words. It is used to adjust the text without any change in the content of the text.

Arabic language has different symbols as diacritical marks(Harakat) which are used to distinguish between words that have the same spelling to represent vowel sounds like: جَمَل or جُمَل . The use of diacritics in text is optional in written standard Arabic language and the well known Arabic readers can read a text without diacritics correctly based on the context. Figure 4 shows the eight main Arabic diacritics.

| | | | |
|---|---|---|---|
| Fatha | ـَ | Kasrah | ـِ |
| Dhammah | ـُ | Sukkon | ـْ |
| Shaddah | ـّ | Tanween Fath | ـً |
| Tanween Kasr | ـٍ | Tanween Dham | ـٌ |

Figure 4. The eight main Arabic diacritics[7].

## 3. ARABIC TEXT WATERMARKING

Many mechanisms and researches have been applied to hide data within texts and in different languages depending on the feature and characters of the letters of each language. Text watermarking methods can be classified into two main classes: linguistic coding and formatting or appearance coding methods. Figure 5 shows the classification of methods that could be applied on Arabic text . Some of these methods are concerned with Arabic language and the others are general.

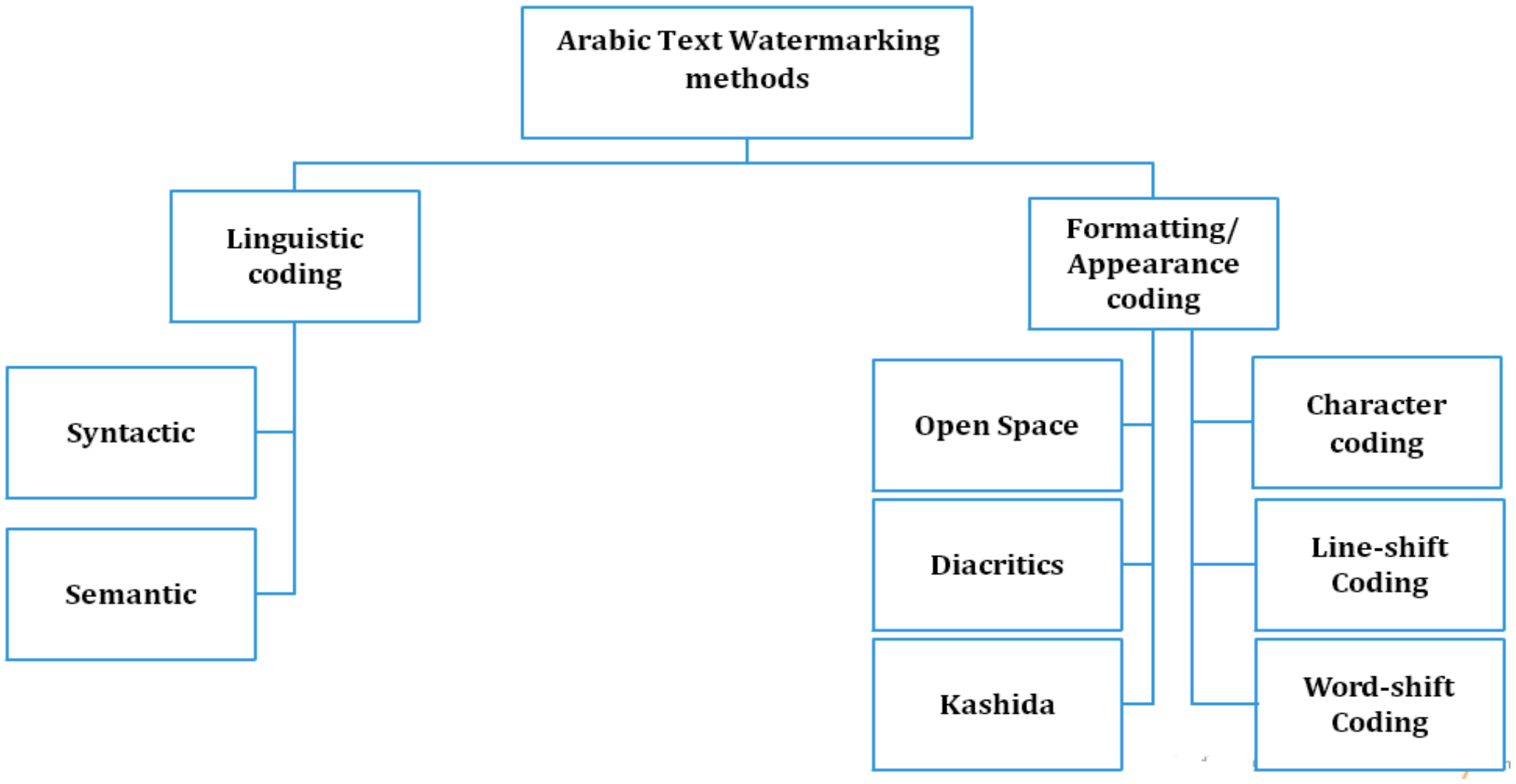

Figure 5. The classification of methods that could be applied on Arabic text.

### 3.1. Linguistic coding

The Linguistic coding approach is based on understanding the main components of the text in specific language .It has two categories .The first one is called syntactic approach which aims to apply some transformations on text structure to embed watermark, such as moving the location of adverbial, adding the form subject, or changing Active to Passive[20].

The second one is a semantic approach where changing is made on text content to insert the watermark, such as replacing certain words with their synonyms[21] , use of abbreviations or changing the order of punctuations[22]. These methods protect the information in case of retyping or using OCR programs. However, they have low capacity and the semantic approach may alter the meaning of the text which is not acceptable for sensitive text like Holly Quran or poems.

## 3.2.Formatting coding

This approach uses the physical formatting of text by making some modifications in the existing text to hide secret information[23].The modifications could be done by adding or deleting spaces or new characters or any change in the appearance of the text. It have the following categories:

### 3.2.1. Open space methods

The process of hiding in this way is done by the addition of extra white spaces in the text .This white space can be placed at the end of each line or at the end of each paragraph, or between words as this method can be applied to any arbitrary text so that does not raise the reader's attention .

In anyway, the size of the hidden information using this method is a little also, some text editors programs automatically delete these white spaces and thus destroy this hidden information [24].It could be applied in any language containing spaces.

### 3.2.2. Diacritics-based methods

The authors in[7] use eight different diacritical symbols to hide binary bits depending on the hidden message in the original Arabic cover media. While the diacritic Fatha has frequency occurrence higher than the other seven diacritics ,the researchers assign a one bit value of 1 to it. The remaining seven diacritics used to represent a value of one bit of 0. The embedded data are then extracted by reading the diacritics from the document and translating them back to binary. In this method ,the omitted diacritic is wasted because it is not hiding anything. Figure 6 shows an example.

| Cover Text | Output Text |
| --- | --- |
| حَدَّثَنَا سُفْيَانُ عَنْ يَحْيَى عَنْ مُحَمَّدِ بْنِ إِبْرَاهِيمَ التَّيْمِيِّ عَنْ<br>عَلْقَمَةَ بْنِ وَقَّاصٍ قَالَ سَمِعْتُ عُمَرَ رَضِيَ اللَّهُ عَنْهُ يَقُولُ<br>سَمِعْتُ رَسُولَ اللَّهِ صَلَّى اللَّهُ عَلَيْهِ وَسَلَّمَ يَقُولُ إِنَّمَا الْأَعْمَالُ<br>بِالنِّيَّةِ وَلِكُلِّ امْرِئٍ مَا نَوَى فَمَنْ كَانَتْ هِجْرَتُهُ إِلَى اللَّهِ عَزَّ<br>وَجَلَّ فَهِجْرَتُهُ إِلَى مَا هَاجَرَ إِلَيْهِ وَمَنْ كَانَتْ هِجْرَتُهُ لِدُنْيَا<br>يُصِيبُهَا أَوْ امْرَأَةٍ يَنْكِحُهَا فَهِجْرَتُهُ إِلَى مَا هَاجَرَ إِلَيْهِ | حَدَّثَنَا سُفْيَان عَن يَحْيَى عن مُحَمَّدِ بْنِ إِبْرَاهِيمَ التَّيْمِي عَنْ<br>علقمة بْنِ وقَّاص قَالَ سمِعت عمر رَضِيَ اللَّهُ عَنْهُ يقول<br>سمعت رَسولَ الله صَلَّى اللَّهُ عَلَيه وَسلَّمَ يقُولُ إنَّما الأعمال<br>بالنيَّة وَلِكُل امرئ مَا نوى فمَنْ كَانَتْ هجرته إلَى الله عزّ<br>وجلّ فهجرته إلَى مَا هَاجر إلَيْهِ ومن كَانَتْ هجرته لدنيا<br>يصيبها أَوْ امرأة ينكحها فهجرته إِلَى ما هاجر إليه |

Figure 6. Example of using Arabic diacritics-based method[7].

The authors in[25] try to put multiple invisible instances of Arabic diacritic marks over each other equal to the binary number representing the message . They use more than one diacritical secret bit at a time. This method has text and image approaches with different scenarios. Figure 7 shows darkening of the black level of the diacritics by multiple instances in the image approach. If the capacity in this method is high ,the robustness of the printing is low and it gives invisible watermarking. However, if the capacity in this method is low the robustness to the printing is high and giving slightly visible watermarking. In both approaches , special Arabic font is used for data hiding process.

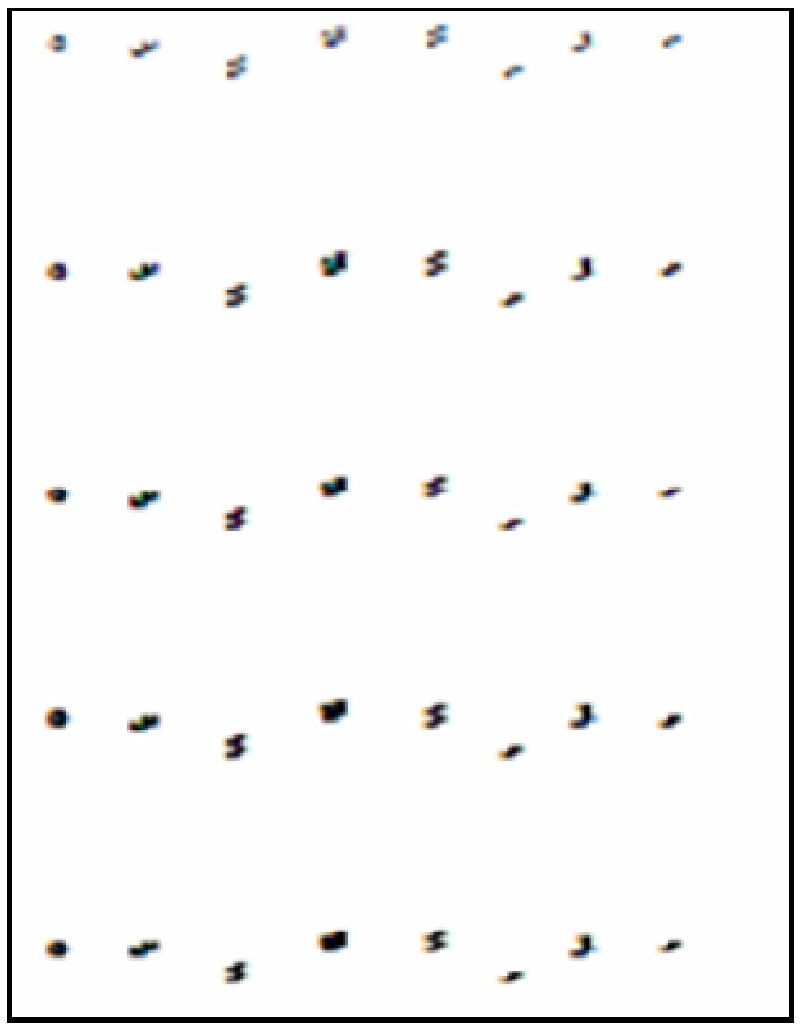

Figure 7. The image of diacritics, from a single instance in first row up to 5 repetitions in the fifth[25].

The main contribution in [8] is to use even omitted diacritics to hide secret bits. If the secret bit is 1 then the diacritic present as it is, but if it is 0 then the diacritic was removed as shown in figure 8. In this method the number of hidden bits is about twice the number of the shown diacritics. That is because the removed diacritics are also hiding bits which means that its capacity as twice as the capacity of first method [7] but it has a weak robustness.

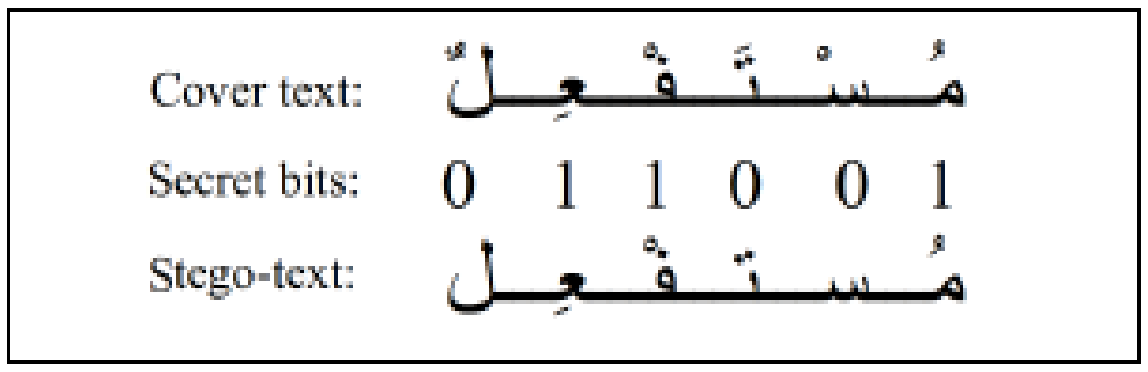

Figure 8. Example of high capacity diacritics-based method[8].

In [26] the authors apply text steganography technique to hide the information by using a reversion for Harakat in the Arabic languages to make their new technique of hiding the Arabic text as in figure 9. The study reverses the original manner of the Fatha from a small line inclining left above the letter to the right by installing new font properties. The regular Fatha is used to encode one and the reverse Fatha is used to encode zero .This method can be applied successfully on printed documents and faced the OCR techniques. But it is a font dependant method and makes visible change in the text. A comparison between some of the diacritics-based methods is shown in Table 2.

| Cover Text | Output Text |
|---|---|
| مُجْتَبَىٰ وَهُوَّ مَحَاضِر | مُجْتَبَىٰ وَهُوَّ مَحَاضِر |

Figure 9. Example of Reverse Fatha method [26].

### 3.2.3 Kashida -based methods

This category depends on the existing of the extension character in Arabic language "Kashida" . It is used in text steganography to hold the secret bits in different algorithms.

The first use of this category was in 2007 where the authors in [27] use letter points and extensions. Their approach utilizes the characteristic of having points within more than half the text letters in Arabic language. They use the pointed letters with extension to hold secret bit 1 and the un-pointed letters with extension to hold secret bit 0 as shown in figure 10 . The proposed steganography method can have the option of adding extensions before or after the letters.

| | |
|---|---|
| Secret bits | 110010 |
| Cover-text | من حسن اسلام المرء تركه مالا يعنيه |
| Steganographic text | مـن حسن اسلام المرء تـركـه مـالا يـعنـيه<br>↑ ↑ ↑ ↑ ↑ ↑<br>1 1 0 0 1 0 |

Figure .10 Hiding secret bits using extension character[27].

The security of the first approach is less since a word may have many inserted Kashidas representing secret bits which reflect the existence of hidden information in the text cover media. For this reason the researchers in[15] propose approach restricts the number of Kashidas used per word to make more confusion to detect the secret information. The main idea of the proposed approach is that a word in a cover text, that have some possible extendable characters, can represent some values within specific range by inserting at most a specific number of Kashidas. This method try to use the locations of possible extendable characters within a given word in the cover text media to hide secret data bits. The secret data is represented within a word by inserting Kashidas after some extendable characters in the word. The researchers propose different scenarios based on the maximum number of Kashida possible to be inserted per word: one, two or three kashidas. This approach gives better capacity and more security than previous method. But it is not robust against retyping process. Figure 11 shows the security improvement of this method comparing with first one.

| | |
|---|---|
| Secret bits | 001010 |
| A word in the cover text | سنمتَعهم |
| Output of proposed method | سـنمتَعـهم |
| Output of the method in [ 14] | سـنـمـتَـعـهم |

Figure 11.Example of Improving security and capacity method using extensions[15].

In [27]while a letter found without Kashida it doesn't represent any secret bit, which is means good capacity but less security. So the anthers in[16] propose approach to use a secret key to generate random Kashida characters added to Arabic e-text words . Kashidas are inserted in the words based on a secret key representing secret bits. The proposed watermarking method made the task of an attack much harder compared to previous similar methods but less capacity than diacritic based methods.

The researchers in[28] develop new algorithm for Arabic text based on the fact that both of the extension letter 'Kashida' and the Zero width character do not have any change in the word meaning when it connected to other letters. The new proposed algorithm used Zero Width and Kashida Letters (ZKS) to hide 2 bits per each connective character as shown in figure 12. This method increases the hidden bit capacity. Figure 13 shows the ZKS method.

| Extension | Zero Width | Code | Letter effect |
|---|---|---|---|
| No | No | 00 | No EFFECT |
| Yes | No | 01 | Extension |
| No | Yes | 10 | Zero width |
| Yes | Yes | 11 | Extension + Width |

Figure 12. ZKS algorithm[28].

| | |
|---|---|
| Cover Object | كان ساحل مصر الشمالي سلة غذاء مصر والامبراطورية<br>الرومانية التي كانت تحتل مصر قبل الإسلام. فقبل بناء<br>السد العالي، اعتمد المصريون على مياه النيل فى<br>الزراعات الصيفية في الوادي والدلتا كما اعتمدوا على مياه<br>الأمطار في زراعة القمح |
| Stego Object | كـان ساحل مـصـر الـشـمـالـي سـلـة غذاء مـصـر<br>والـامبراطـورية الرومـانـيـة التي كانت تـحـتـل مـصـر<br>قـبـل الإسـلـام. فقبل بـناء الـسـد الـعـالـي، اعتمد<br>الـمصريـون عـلـى مـيـاه الـنيل فـى الزراعات الـصـيـفـيـة<br>فـي الـوادي والدلـتـا كـمـا اعـتـمـدوا عـلـى مياه الأمـطـار<br>في زراعة الـقـمـح |
| Hidden Bits | 1001001000001110101100010111101100011111<br>110011101111100111101000001001011100011<br>00000001011010001111111000000101001 |

Figure 13. Example using ZKS algorithm [28].

The authors in [29] utilize the Kashida by encoding the original text document with Kashida according to a specific key which will be produced before the encoding process .Kashidas are inserted before a specific list of characters {أ، إ، آ، ا، د، ذ، ر، ز، و، ؤ }until the end of the key is reached where the kashida is inserted for a bit 1 and omitted for a bit 0 as shown in figure 14. This process repeated until the end of the document is reached in a round robin fashion. The proposed technique proved to achieve document protection and authenticity giving high robustness than other Kashida methods but it provide less capacity.

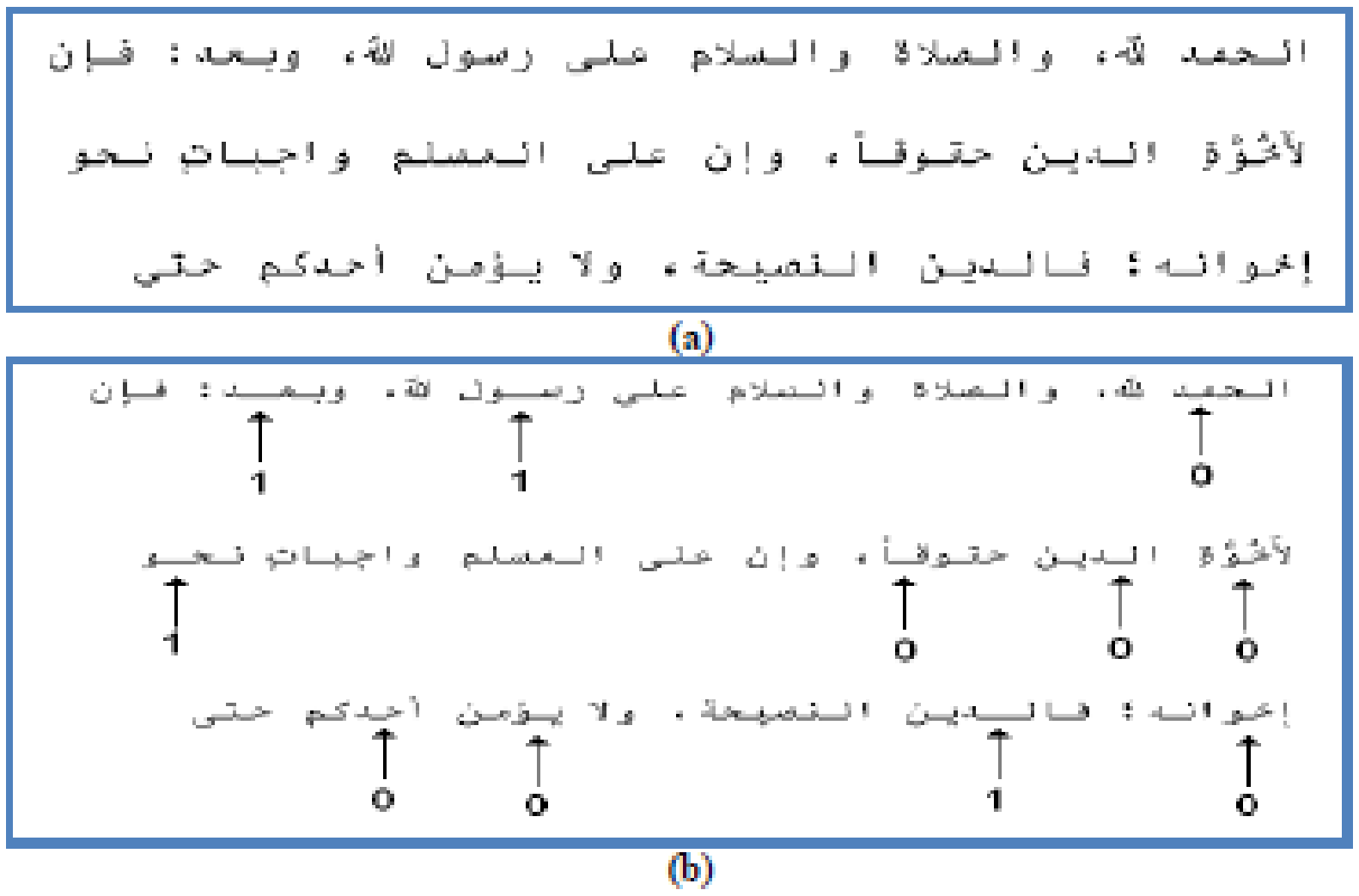

Figure 14. a) Original Text with Key = 00101000110, and b) Watermarked text[29].

The researches in[17] use the same algorithm developed in [29] with different sets based on character frequency recurrence properties. They divide the Arabic letters into two sets: set A containing the first 14 letters with higher frequencies and set B containing the reminder letters. They propose two methods (Method-A and Method-B). In Method-A replace the set of characters in[29] with set A. In Method-B, the kashida is inserted before the character in two cases: if the key bit is zero and the character in Set-A and if the key bit is one and the character in Set B. Figure 15 shows an example. These methods give higher capacity than [29]and improved imperceptibility as compared to the other Kashida based methods. Table 3 shows comparison between some of the kashida-based methods.

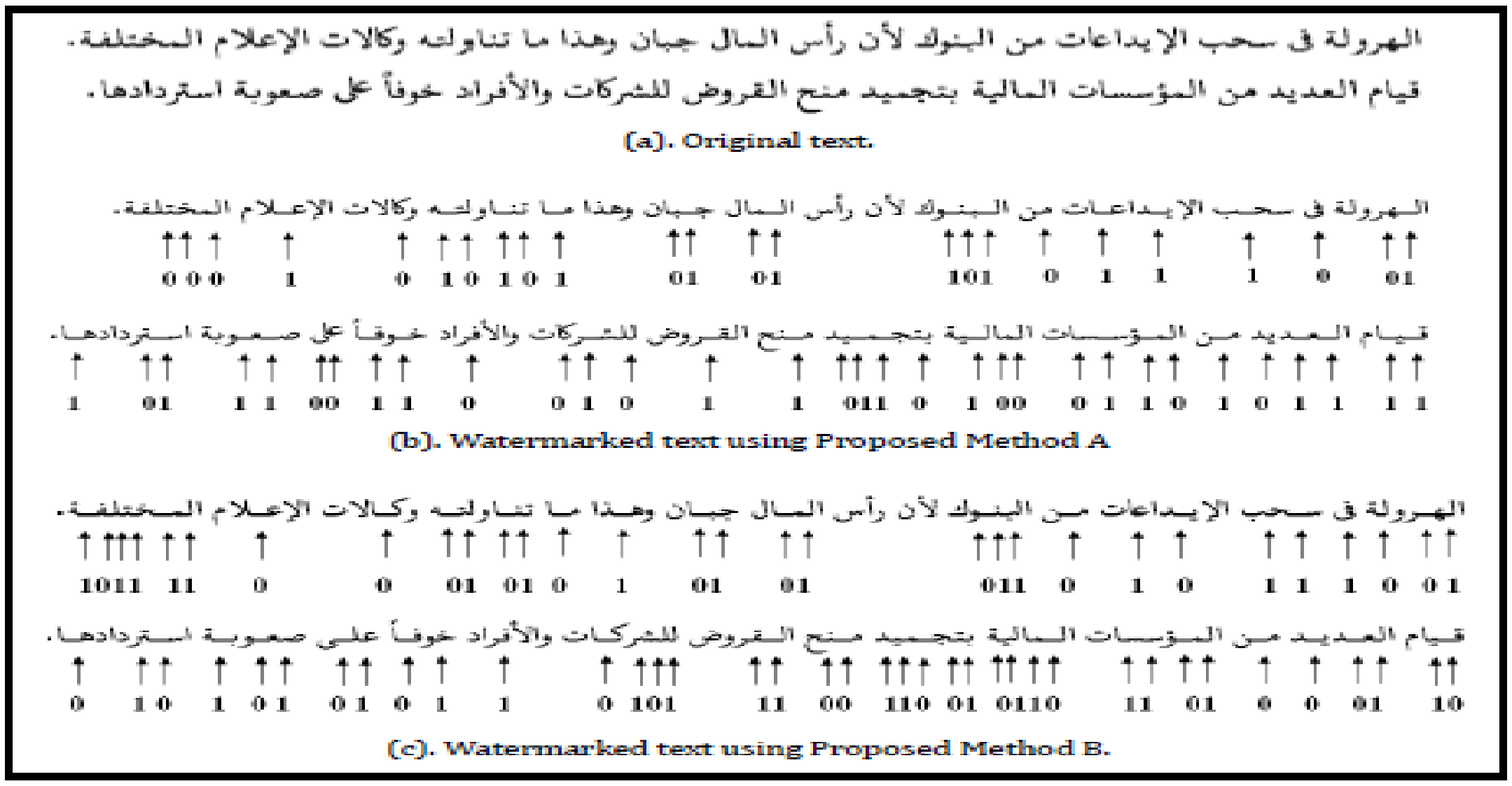

Figure 15. An example of a sampled text after applying the watermark key[17].

### 3.2.4. Line-Shift Coding

In line shifting method[30], the lines are moved vertically upward or downward to hide information. It can be applied to images without the need for original image because of the regular spaces between the lines in the image.

Lines displacement process is the most noticeable by the reader, but it has a good strength and resistance to noise due to the length of the lines so it is more suitable for use with images .It has low capacity where it can hold one bit per line. Figure 16 shows an example of this method.

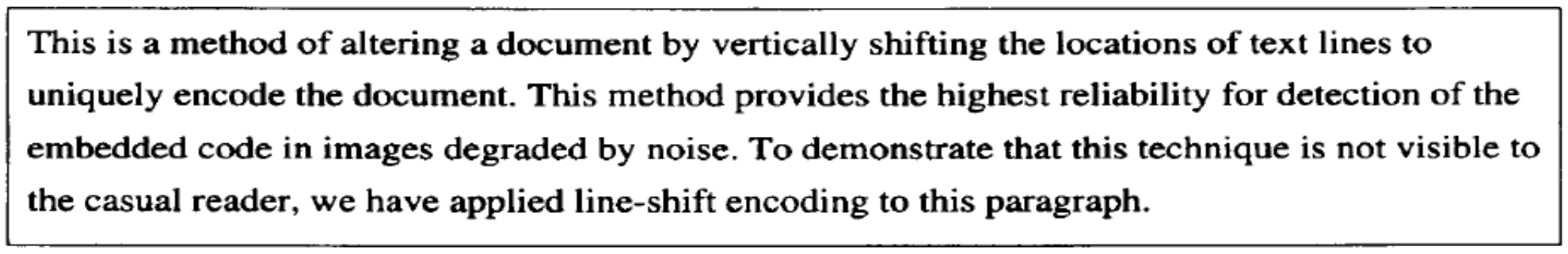
This is a method of altering a document by vertically shifting the locations of text lines to uniquely encode the document. This method provides the highest reliability for detection of the embedded code in images degraded by noise. To demonstrate that this technique is not visible to the casual reader, we have applied line-shift encoding to this paragraph.

Figure 16. Example of line-shift coding. The second line has been shifted up by 1/300 inch[30].

### 3.2.5. Word-Shift Coding

In word shifting method[30], the words are moved horizontally making spaces expanding between words to hide information. This method needs the presence of the original file or the original image due to the possibility that the file contains a variable number of spaces between adjacent words.

It is better and more effective for use with files because they are less attractive to the reader's attention. It's capacity is about on bit per word. Figure 17 shows an example of this method.

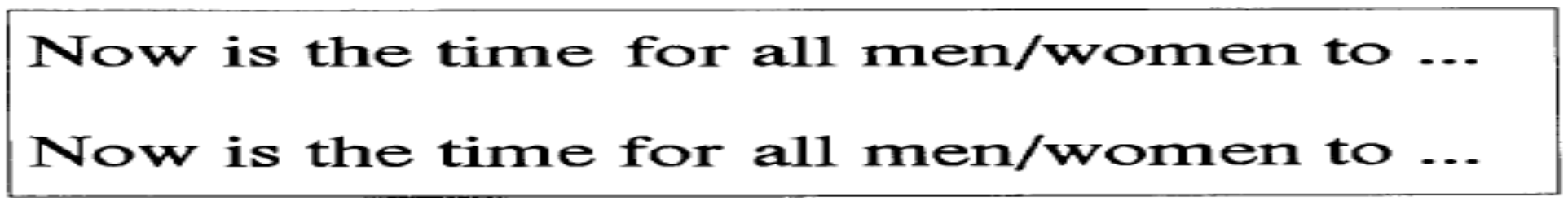
Now is the time for all men/women to ...

Now is the time for all men/women to ...

Figure 17. Example of word-shift coding. The top text line has added spacing before the "for" [30].

### 3.2.6. Feature Coding

This approach modifies features a little bit to embed a watermark. In [30] the length of end lines of letters, b, d, h, etc is altered by extending or shortening it. Line and word shifting could be applied in any language even Arabic but the feature coding is done on a language based on its features. All of these methods could be applied to a format file or to a binary image of a document. We will discuss some of studies based on feature coding in the Arabic language .
The authors in[9] proposed an approach depends on the characteristic of existing of points on the majority of Arabic, Urdu and. Persian letters. These points are used to hide secret binary information. If the secret bit equals to one ,the points within pointed letter is shifted a little upward. If the secret bit equals to zero , the points' location does not change. This approach has high capacity where the Arabic language has 15 pointed letters from its all letters(28). It is also utilized with extension giving better results in[14]. The recovery process is acceptable even in the case of printed text. However , the retyping process will remove all the secret hidden bits. Also, this method requires creating of a special font .Figure 18 shows vertical displacement of the points for the letter NOON.

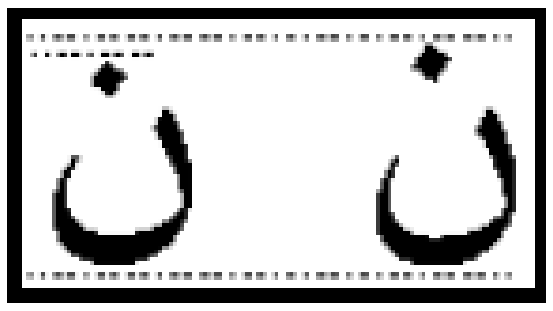

Figure 18. Vertical displacement of the points for the letter NOON[9].

The author proposed a method for Steganography in Persian and Arabic texts in[10] based on feature coding using La" word. This word is obtained by connecting "Lam" and "Alef" letters in single word. The hiding process is done based on the existing of two forms of this word: special form "La" (" لا") which has a unique code and normal form "La" ("لـا") by inserting Arabic extension character between "Lam" and "Alef" letters. For hiding bit zero, they use the normal form of La, whereas bit one is hidden using the special word. This method is not limited to electronic documents and could be used on printed documents. . However, this method has low capacity since the using of "La" word is limited in Arabic text. Also, it increases the file size and gives abnormal appearance of the text. For this reasons the authors in[11] proposed an improved version of the previous method to solve these problems. They use different Unicode of "Lam" and "Alef" to form "La" word in both spatial and normal form based on the fact that each letter can have four different shapes depends on its position in the word. The improved method does not change the file size and gives normal appearance of the text with the same low capacity ratio. Figure 19 shows the difference between these methods.

| Watermarking bits | 1001 |
| --- | --- |
| Cover-text | کلا لازم است قبل از وقوع بلایای طبیعی اقدامات ایمنی لازم انجام شود. |
| Original "LA" method | کلا لـازم است قبل از وقوع بلـایای طبیعی اقدامات ایمنی لازم انجام شود.<br>↑ ↑ ↑↑<br>1 0 0 1 |
| Our Improved method | کلا لازم است قبل از وقوع بلایای طبیعی اقدامات ایمنی لازم انجام شود.<br>↑ ↑ ↑↑<br>1 0 0 1 |

Figure 19. Original "LA" Steganography and improved method example [11]

Also he proposed a new method for Steganography in Persian and Arabic texts in[13] based on existing of pseudo space in some Persian words. He insert the normal space after pseudo-space to hide the bit 1 and no change in pseudo-space to hide the bit 0 .The pseudo-space is coded in the Unicode hex notation as 200C.It is a small space used to separate the same word into two parts.
A new steganography method for Persian and Arabic texts is presented in[12]based on the existing of two similar shape letters "Ya" « ی» and "Kaf" « ک» with different Unicode. The authors use the Persian characters of « ک» or « ی» to hide the bit 0 and the Arabic characters of « ك» or « ي» to hide the bit 1 . This method has advantage of high imperceptibility since no changing occurs on the appearance of the text. However, it has low capacity based on using these two letters on the text. Table 4 shows comparison between some of Arabic watermarking methods based on feature coding.

## 4. CONCLUSIONS

While the Arabic text is the most extensively used than other kind of media, it will need a good method to protect its privacy. Some of Arabic texts watermarking methods are described within a specific classification. The comparison tables which summarize these methods with their advantages and disadvantages are presented. Most of diacritics-based methods are simple to implement and give higher capacity and robustness than other. However, it cannot be used in text in which the appearance of all diacritics is important like Holly Quran and most e-text today is written without diacritics. Kashida-based methods have good capacity and could be used in printing document with different font formats. However, they are easily to detect or observed. So, many researchers add more security features to decrease the number of kashidas and enhance the imperceptibility. The methods based on the Unicode give higher imperceptibility where the watermark is invisible but they restricted on electronic text only. Even of the simplicity of shifting (line, word or points) methods, it have a big drawback which is the ability to destroy the watermark is very high when retyping or printing and they are noticeable by OCR.

Table 2. Comparison between some of diacritics-based methods.

| Author(s) | The used features for hiding the watermark | | Robustness | | | | | Capacity | | Visibility | Evaluation resources |
|---|---|---|---|---|---|---|---|---|---|---|---|
| | | | Printing | OCR | Copying & pasting | Font changing | Retyping | % | bit/KB | | |
| M. Aabed, S. Awaideh, A. Elshafei and A. Gutub [7] | Eight different diacritical symbols. | | ✓ | ✓ | ✓ | ✓ | ✓ | 3.28 | 269 | Slightly visible | A part of Musnad Al-Emam Ahmed online book contains 7305490 characters and 1037265 words. |
| M.L. Bensaad, and M.B.Yagoubi [8] | Use even omitted diacritics for hidden. | | ✓ | ✓ | ✓ | ✓ | ✓ | 6.44 | 528 | Slightly Visible | 5 large text files of different sizes taken from arbitrarily chosen Arabic books that are almost fully diacritized. |
| A. Gutub Y. S. Elarian, S. M. Awaideh, and A. K. Alvi [25] | Multiple invisible instances of Arabic diacritic. | Text + soft copy. | ✗ | Not specified | Not specified | Not specified | Not specified | High | | Invisible in code | Not specified |
| | | Image + soft copy. | ✓ | | | | | Very low | | Slightly visible | |
| | | Image + hard copy. | ✓ | | | | | Moderate | | Slightly visible | |
| A. Shah, and M. S. Memon [19] | Using a reversion of Harakat(Fatha). | | ✓ | ✓ | Not specified | ✗ | Not specified | Not specified | Not specified | Slightly Visible | Not specified |

Table 3.Comparision between some of kashida-based methods.

| Author(s) | The used features for hiding the watermark | | Robustness | | | | | Capacity | | Visibilit y | Evaluatio n resources |
|---|---|---|---|---|---|---|---|---|---|---|---|
| | | | Printing | OCR | Copying & pasting | Font changing | Retyping | % | bit/ KB | | |
| A. Gutub, L. Ghouti, A. Amin, T. Alkharobi and M. Ibrahim [14] | Letter points and extensions. | | ✓ | ✗ | ✓ | ✓ | ✗ | 1.22 | 100. 32 | Visible | The corpus of Contempor ary Arabic is reported to have 842,684 words from 415 diverse texts from websites. |
| F.,Al-Haidari, A. Gutub , K. Al-Kahsah and J. Hamodi [15] | Restricting the number of Kashidas used per word. | | ✓ | ✗ | ✓ | ✓ | ✗ | 3.16 | 260 | Visible | 4 Different cover text media sizes |
| A. Gutub, F. Al-Haidari, K. M. Al-Kahsah and J. Hamodi [16] | A secret key to generate random Kashida characters. | | ✓ | ✗ | ✓ | ✓ | ✗ | 2.8 | 229 | Visible | E-text cover files with different sizes |
| A. Odeh and K. Elleithy [28] | Using Kashida with the Zero width letter. | | Not specified | Not specified | Not specified | Not specified | Not specified | Not specified | Not specified | Visible | Not specified |
| Y. Alginahi, M. N. Kabir and O. Tayan [29] | Kashidas are inserted before a specific list of characters ( أ، ا، ز، ر، ذ، د، ئ آ، إ، د و، ) | | ✓ | ✗ | ✓ | ✓ | ✗ | 0.00 02 | 16.8 8 | Visible | 4 Different documents with different lengths. |
| Y.Alginahi , M. Kabir and O. Tayan [17] | Kashidas are inserted before two sets based on character frequency | Method A | ✓ | ✗ | ✓ | ✓ | ✗ | 0.00 08 | 71.1 8 | Visible | Four different documents with different lengths |
| | | Method B | ✓ | ✗ | ✓ | ✓ | ✗ | 0.0 038 | 78.7 1 | | |

Table 4. Comparison between some of Arabic watermarking methods based on feature coding.

| Author(s) | The used features for hiding the watermark | Robustness | | | | | Capacity | | Visibility | Evaluation resources |
|---|---|---|---|---|---|---|---|---|---|---|
| | | Printing | OCR | Copying & | Font changing | Retyping | % | bit/ KB | | |
| M. H Shirali-Shahreza, and M. Shirali-Shahreza [9] | Points shifting. | ✓ | ✗ | ✗ | ✗ | ✗ | 1.37 | 112.8 | Slightly Visible. | Sport pages of 10 Iranian newspapers. |
| M. Shirali-Shahreza [10] | Using "La" word with extension. | ✓ | ✗ | ✓ | ✓ | ✗ | 0.012 | 1.17 | Visible. | Sport pages of 10 Iranian newspapers. |
| M. H Shirali-Shahreza, and M. Shirali-Shahreza [11] | Using "La" word with different shapes | ✓ | ✗ | ✓ | ✓ | ✗ | 0.012 | 1.17 | Visible | Sport pages of 10 Iranian newspapers |
| M. H Shirali-Shahreza, and M. Shirali-Shahreza [12] | Existing of two similar shape letters “Ya” « ى» and “Kaf” « ک» with different Unicode. | ✗ | ✗ | ✓ | ✓ | ✗ | 0.40 | 33.25 | Invisible. | Sport pages of 8 Iranian newspapers. |
| M. Shirali-Shahreza [13] | Using pseudo-space. | ✗ | ✗ | ✓ | ✓ | ✗ | 0.05 | 4.18 | Invisible. | Sport pages of 2 Iranian newspapers. |

## Authors

Reem Ahmed Alotaibi holds a BA degree in Computer Science with Honours, from Taif University (TU) in 2010. She worked as Teaching Assistant from 2011 until 2013 at Faculty of Computer Science and Information Technology at Taif University . Now she is working in her Master Degree at King Abdulaziz University Jeddah, Saudi Arabia . She is interested in the areas of data hiding and pattern recognition.

Lamiaa A. Elrefaei received her B.Sc. degree with honors in Electrical Engineering (Electronics & Telecommunications) in 1997. The M.Sc. in 2003 and Ph.D. in 2008 both in Electrical Engineering (Electronics) from Faculty of Engineering at Shoubra, Benha University, Egypt. She has held a number of faculty positions at Benha University, as Teaching Assistant from 1998 to 2003, as an Assistant Lecturer from 2003 to 2008, and as an Assistant Professor (referred to as Lecturer position in the Egyptian academic system) from 2008 to date. She is currently serving as an Assistant Professor at King Abdulaziz university, Jeddah, Saudi Arabia. She is a member of IEEE. Her research interests include computer vision, information security, and computational intelligence.